# A common warming-dependent decline emerges from divergent projections of tropical cyclone frequency

Cong Gao[1*], Ning Lin[1*]

Affiliations:

[1]Department of Civil and Environmental Engineering, Princeton University, Princeton, NJ, USA

*Corresponding authors. Emails: cong.gao@princeton.edu; nlin@princeton.edu

**How global tropical cyclone frequency will respond to climate warming remains uncertain, with projections spanning both increases and decreases[1–7]. Here we use an environment-dependent probabilistic tropical cyclone model driven by 14 CMIP6 models under intermediate- and high-emissions scenarios to identify a common response across these divergent projections. Initially weak global changes evolve into a clearer late-century decline, driven by increasingly negative changes in the Southern Hemisphere and diminishing positive changes in the Northern Hemisphere. The long-term decline is stronger under higher emissions. Counterfactual decomposition shows that this stronger decline cannot be explained by the sum of the individual predictor contributions, which largely offset one another, but is instead dominated by negative non-additive joint effects among absolute vorticity, vertical wind shear, relative humidity and potential intensity. Across 28 model–scenario projections, the global tropical cyclone frequency trend scales linearly with the level of global warming. Applying the fitted warming–frequency relationship at 2 °C of warming relative to 1980–2014 narrows the 95% confidence interval for the projected frequency change from [−21.9%, +10.3%] to [−9.1%, −2.5%]. These results reconcile divergent projections through a common warming-dependent decline and identify the joint reorganization of large-scale environmental conditions as its primary driver.**

## INTRODUCTION

Tropical cyclones are among the most consequential climate hazards, yet how their global frequency will respond to anthropogenic warming remains unresolved[1–7]. A warmer climate alters the large-scale environmental conditions that regulate tropical cyclone genesis through both thermodynamic and dynamical changes[8–12]. However, these changes do not point uniformly toward either more or fewer storms, with projections of global tropical cyclone genesis frequency differing across climate models and generations of the Coupled Model Intercomparison Project (CMIP)[4,6,7,13–15]. Although most studies have projected declines in global tropical cyclone genesis frequency, some have projected increases[6,16–20], leaving the sign of the long-term response under continued debate. Analyses of the historical record using different methods add further nuance: tropical cyclone detection and tracking in reanalyses indicate an approximately 13% decline in global tropical cyclone frequency over the twentieth century relative to pre-industrial levels[21], whereas statistical–dynamical downscaling of reanalyses shows only weak or insignificant trends across most global metrics[22]. This continued uncertainty is reflected in the Sixth Assessment Report (AR6) of the Intergovernmental Panel on Climate Change (IPCC), which found medium confidence that global tropical cyclone genesis frequency will decrease or show little change with future warming[5]. Reducing this uncertainty is essential for reliable projections of how tropical cyclone climatology will evolve in a warmer climate, and the large inter-model spread in projected storm occurrence points to the need to identify the sources of divergent projections[4].

Previous studies have largely focused on whether global tropical cyclone frequency increases or decreases under global warming[4,7], but have offered less insight into why projections differ across climate models. Variations in model resolution[23], storm detection and tracking methods[18,24], convective and microphysical parameterizations[25,26], and air–sea coupling and upper-ocean representation[27] can all influence projected storm occurrence, making it difficult to distinguish methodological differences from physically meaningful signals. Moreover, most studies have emphasized the net change in frequency itself, rather than asking whether models with stronger warming also exhibit systematically larger declines. This leaves unresolved the question of whether inter-model spread mainly reflects noise and structural uncertainty, or whether it contains an emergent warming-dependent signal tied to changes in the environmental controls on tropical cyclone genesis.

Genesis potential indices, pioneered by Gray[12] and developed further through subsequent formulations[10,11,28–30], estimate tropical cyclone genesis frequency from the large-scale environmental conditions simulated by climate models. Their key advantage is that they diagnose storm occurrence from the environmental controls on genesis, rather than from explicitly simulated tropical cyclones, which remain under-resolved in most climate models[31]. However, projections based on these previously developed genesis potential indices differ substantially from those of explicitly tracked storms and do not reproduce the decrease in tropical cyclone numbers projected by most climate models[13]. More comprehensive statistical or statistical-deterministic downscaling methods have been proposed[32–34], including the recently developed Princeton Environment-dependent Probabilistic Tropical Cyclone Global (PepC-Global)[35] model, which advances the PepC[34] model to the global scale, using the large-scale environment to generate synthetic tropical cyclones over global ocean basins.

Here we use the PepC-Global genesis module, a support vector classification-based Poisson process framework driven by absolute vorticity, vertical wind shear, relative humidity and potential intensity (see Methods)[35], to examine projected changes in global tropical cyclone genesis frequency and to test whether the spread across climate models and emission scenarios contains a coherent warming-dependent signal. We select 14 CMIP6 models under Shared Socioeconomic Pathways SSP245 (an intermediate-emissions scenario) and SSP585 (a high-emissions scenario)[36] to span a broad range of projected global warming levels (Extended Data Fig. 1 and Extended Data Table 1).

We show that a clearer global decline in tropical cyclone frequency emerges by late century under stronger forcing, with a stronger long-term decline under SSP585 than under SSP245. We further show that this stronger decline cannot be explained by the sum of the individual predictor contributions, but instead stems from substantial non-additive joint effects among the environmental controls on genesis. The long-term trend in global tropical cyclone frequency scales linearly with the global warming level, at −0.0303% per year per °C. The warming–frequency scaling we identify substantially reduces the existing uncertainty in the projection of frequency change over the century, narrowing the 95% confidence interval from [−21.9%, +10.3%] to [−9.1%, −2.5%] for 2 °C of warming, reconciling the divergent projections through a common forced response.

## RESULTS

### Emergence of a late-century decline in tropical cyclone frequency

Following previous projection studies, including Knutson et al. (2020)[4], we express future changes in tropical cyclone frequency, which is obtained from the PepC-Global genesis module (see Methods), as percentage changes relative to present-day (1980–2014) conditions to facilitate comparison across models with different baseline climatologies. Averaging model-specific percentage changes gives each model equal weight in the relative response, whereas a multi-model mean calculated directly from frequencies would be more strongly influenced by models with higher baseline climatological frequencies. We select 1980–2014 as the reference period because it falls within the modern satellite era, when improved global storm detection provides more reliable tropical cyclone frequency records than those before 1980. In the near term, increasing tropical cyclone frequency in the Northern Hemisphere and decreasing tropical cyclone frequency in the Southern Hemisphere combine to yield very similar global responses under SSP245 and SSP585, with both projecting slight increases in global tropical cyclone frequency (Fig. 1a). By mid-century, the hemispheric responses remain qualitatively similar, but the Northern Hemisphere increase weakens and the Southern Hemisphere decline strengthens (Fig. 1b). These changes in the two hemispheres shift the global response downward from the near term to mid-century, yielding a near-zero change in global tropical cyclone frequency under SSP245 but a weak decline under SSP585, both with substantial uncertainty. By late century, the Southern Hemisphere decline deepens further, and the Northern Hemisphere increase continues to diminish, leading to a more pronounced decline in global tropical cyclone frequency, particularly under SSP585 (Fig. 1c). The stronger late-century decline in the Southern Hemisphere is broadly consistent with previous studies[4,37,38].

This late-century decline is robust, appearing in the multi-model mean and supported by agreement on the sign of the change across the 14 CMIP6 models (Extended Data Fig. 2a–f). In the near term, there is no consensus on the sign of global tropical cyclone frequency change under either SSP245 or SSP585 (Extended Data Fig. 2a,b), despite the weak positive multi-model mean signal (Fig. 1a). By contrast, negative changes are already more consistent in the Southern Hemisphere, with 10 of 14 models projecting declines under SSP245 and 13 of 14 under SSP585, corresponding to likely and very likely declines, respectively, on the IPCC likelihood scale. By mid-century, the global response shifts further downward: consensus remains absent under SSP245 (Extended Data Fig. 2c), whereas a likely negative consensus emerges under SSP585 (Extended Data Fig. 2d). This downward shift is accompanied by stronger Southern Hemisphere agreement, with 12 of 14 models projecting declines under SSP245, and by an increase in the number of Northern Hemisphere models projecting declines under SSP585, from 4 of 14 in the near term to 9 of 14 by mid-century. By late century, more models project negative than positive global changes under both SSP245 and SSP585, with 9 of 14 models showing declines under SSP245 and 10 of 14 under SSP585; only SSP585 reaches likely negative agreement, consistent with the more clearly negative multi-model mean response in Fig. 1c. The Southern Hemisphere shows more consistent negative changes than the Northern Hemisphere throughout, supporting its dominant role in the emergence of the late-century global decline.

**Stronger long-term declines under stronger forcing**

Consistent with the clearer late-century decline under SSP585 in Fig. 1c, the late-century scenario contrast is evident in the multi-model mean difference shown in Fig. 1d and is also supported by the model-by-model comparison in Extended Data Fig. 2g. Examination of the annual time series confirms that this contrast reflects a persistent long-term divergence between the two scenarios rather than a feature confined to the late-century window. Linear trends are widely used to summarize long-term climate change[5]. For the global multi-model-mean time series in Fig. 2a, quadratic fits reduce the root-mean-square error relative to linear fits by only 3.17% under SSP245 (4.6775 to 4.5291) and 3.06% under SSP585 (4.3086 to 4.1767). These modest improvements support using linear trends as a parsimonious summary of the 2015–2099 changes. At the global scale, tropical cyclone frequency declines significantly under both scenarios over 2015–2099, with a steeper trend under SSP585 than under SSP245, at −0.1410% year$^{-1}$ and

−0.0649% year$^{-1}$, respectively, with $p < 0.001$ for SSP585 and $p = 0.002$ for SSP245 (Fig. 2a). This stronger global decline under SSP585 is accompanied by contrasting hemispheric responses: in the Northern Hemisphere, the weak decline under SSP245 is not significant, whereas SSP585 shows a significant negative trend, at −0.1018% year$^{-1}$ with $p < 0.001$; in the Southern Hemisphere, both scenarios exhibit significant declines, but the trend is stronger under SSP585 than under SSP245, at −0.1773% year$^{-1}$ and −0.1246% year$^{-1}$, respectively, with $p < 0.001$ for both (Fig. 2b,c). Temporal autocorrelation does not materially affect the results because it is weak, and accounting for it leaves trend significance unchanged while altering slope magnitudes by only 0.13–1.38%. Together, these results show that stronger forcing is associated with a stronger long-term decline in tropical cyclone frequency.

To understand why the long-term decline is stronger under SSP585 than under SSP245, we next decompose the difference in tropical cyclone frequency trend between the two scenarios into individual predictor contributions and non-additive joint effects (see Methods). The total trend difference is negative for the globe, Northern Hemisphere and Southern Hemisphere, indicating a more negative tropical cyclone frequency trend under stronger forcing, and is significant for the globe and Northern Hemisphere but not for the Southern Hemisphere (Fig. 3). Individual predictor contributions are mixed in sign: absolute vorticity at 850 hPa and vertical wind shear contribute negatively, whereas potential intensity contribute positively. A negative contribution from absolute vorticity has also been identified using two genesis potential indices[13]. A positive contribution from potential intensity has also been identified in previous studies[13,18]. However, these individual predictor contributions largely cancel and do not explain the total trend difference on their own, while the contribution from non-additive joint effects is negative and comparable to the total trend in magnitude in all three domains (Fig. 3). Consistent with our results, Emanuel[18] found in a 1% year$^{-1}$ $CO_2$ experiment that non-additive joint effects in the genesis potential index make a negative contribution. In that study, the negative contribution was inferred from the sum of the individual predictors' trends exceeding the trend in the genesis potential index. These results show that the stronger long-term decline under SSP585 arises from substantial non-additive joint effects rather than the sum of the individual predictor contributions.

Tropical cyclone genesis is an ingredient-limited[12], threshold-like process[29], so non-additive joint effects are physically expected because the response to a change in one predictor depends on the states of the others. A favourable shift in one predictor has limited effect unless other ingredients also align, whereas an unfavourable shift in one predictor can strongly inhibit genesis even if other ingredients become more favourable. This asymmetric sensitivity may also help explain why the contribution from non-additive joint effects is negative. The negative contribution from non-additive joint effects can be attributed mainly to unfavourable shifts in low-level absolute vorticity and vertical wind shear. The future change is governed primarily by how environmental conditions combine to alter the frequency of genesis-permitting environments, rather than by the effect of any single factor alone. We therefore caution against attributing the projected change to one or two ingredients in isolation, without accounting for their joint effects with the remaining environmental controls. The projected decline in genesis has likely emerged from a reorganization of the joint environmental state space in the warming climate simulation.

**Warming–frequency scaling of tropical cyclones**

Having shown that tropical cyclone frequency declines more strongly under SSP585 than under SSP245, we finally ask whether this scenario contrast reflects a broader warming-dependent relationship across model projections. Across the 28 CMIP6 model–scenario projections, we use a zero-intercept regression to represent the 2015–2099 tropical cyclone frequency trend as proportional to global warming, assuming that this component vanishes at zero warming. The fitted slope is −0.0339% $year^{-1}$ °$C^{-1}$ ($p = 0.001$; Fig. 4). The 28 projections may not be independent because each model contributes both SSP245 and SSP585 simulations; we therefore assess uncertainty in the zero-intercept regression by combining residual contributions within each of the 14 models and using 13 degrees of freedom for inference. Accounting for this pairing leaves the slope unchanged at −0.03394% $year^{-1}$ °$C^{-1}$, but increases its standard error from 0.00948 to 0.01396% $year^{-1}$ °$C^{-1}$ and changes the p value from 0.00133 to 0.03024. The estimated mean change at 2 °C over 85 years remains −5.8%, with the 95% confidence interval slightly widening from [−9.1%, −2.5%] to [−10.9%, −0.6%]. The zero-intercept regression from Fig. 4 provides a warming-based scaling that can be used to constrain projected late-century changes in global tropical cyclone frequency.

Across the 28 unadjusted CMIP6 model–scenario projections, the mean projected change in global tropical cyclone frequency over the 85-year period 2015–2099 is −8.8%, with a Gaussian 95% model range of [−35.6%, +18.1%]. This range is comparable to the large uncertainty reported in the multi-model assessment of Knutson et al. (2020)[4], which found a median global tropical cyclone frequency change of −14%, with a range from −28% to +22%, for 2 °C of warming. Expressing our raw projections on a common 2 °C warming basis, by scaling each model's projected frequency change by its corresponding global-mean warming, gives a mean change of −5.8% and narrows the Gaussian 95% model range to [−21.9%, +10.3%], but still leaves substantial cross-model uncertainty. In contrast, applying the emergent cross-model warming–frequency scaling directly to the raw projections yields a best-estimate decline of −5.8%, with a narrow 95% confidence interval [−9.1%, −2.5%]. The cross-model relationship thus generalizes the SSP585–SSP245 contrast into a quantitative scaling between warming level and tropical cyclone frequency decline, substantially reducing the projection uncertainty.

## DISCUSSION

Across 28 CMIP6 model–scenario projections, changes in global tropical cyclone frequency remain divergent, spanning both increases and decreases. Nevertheless, a negative warming–frequency relationship emerges across the model–scenario projections. The central result is therefore not that every model projects fewer tropical cyclones, but that the common component of the response becomes increasingly negative with warming. This warming-dependent relationship is consistent with the stronger long-term decline under SSP585 than under SSP245 and yields an expected global frequency change of −5.8% at 2 °C of warming, with a 95% confidence interval of [−9.1%, −2.5%], compared with [−21.9%, +10.3%] before applying the constraint. The global response reflects a pronounced hemispheric asymmetry, arising from a strengthening decline in the Southern Hemisphere and a diminishing increase in the Northern Hemisphere. Importantly, fewer tropical cyclones globally do not imply a commensurate reduction in tropical cyclone risk, which will also depend on changes in storm intensity, rainfall, storm surge, landfall characteristics, and societal exposure.

The decomposition analysis suggests that the projected change in tropical cyclone genesis frequency is governed mainly by non-additive joint effects among large-scale environmental predictors. Tropical cyclone genesis is fundamentally constrained by the simultaneous availability of necessary environmental ingredients and exhibits threshold-like behaviour. The dominance of the non-additive joint effects therefore indicates that the response to climate forcing cannot be explained by environmental changes considered separately. Decreases in low-level absolute vorticity and increases in vertical wind shear contribute importantly to this negative non-additive joint effect. More broadly, these results highlight changes in the large-scale circulation as a central driver of future changes in tropical cyclone genesis frequency.

Several directions could further reduce the uncertainty in projection of how tropical cyclone frequency responds to climate forcing. Here, we did not quantify how much of the projected change arises from internal variability versus the anthropogenic forced response, although most of the projected changes should arise from the forced response, particularly under SSP585. Detection and Attribution Model Intercomparison Project (DAMIP)[39] could help narrow this uncertainty by isolating the forced signal more explicitly, comparing single-forcing DAMIP simulations (e.g., SSP245-GHG and SSP245-aer) with the full-forcing SSP245 projections. Single-model large ensembles, such as the CESM2 Large Ensemble Community Project[40], could also reduce uncertainty by separating the forced signal from internal variability, with the influence of internal variability reduced by averaging across many ensemble members.

## MAIN REFERENCES

## FIGURES

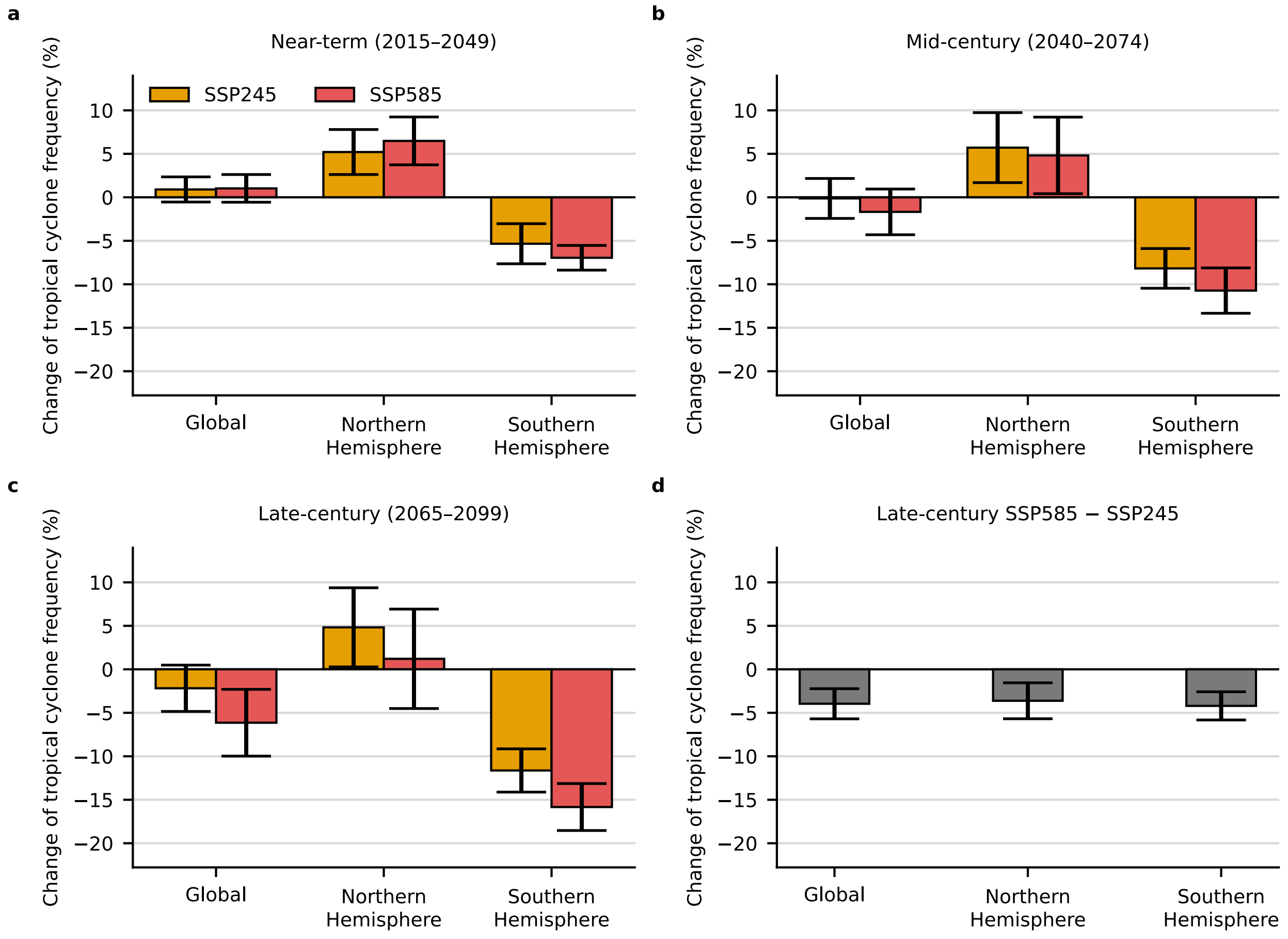

**Fig. 1 | Multi-model mean changes in tropical cyclone frequency under SSP245 and SSP585 for the near-term, mid-century and late-century periods. a**, Near term (2015–2049). **b**, Mid-century (2040–2074). **c**, Late century (2065–2099). **d**, Difference between the late-century multi-model-mean changes under SSP585 and SSP245. Changes are relative to present-day (1980–2014) climatology and are shown for the globe, Northern Hemisphere and Southern Hemisphere as multi-model means across the 14 CMIP6 models. Error bars denote the standard error of the mean across models.

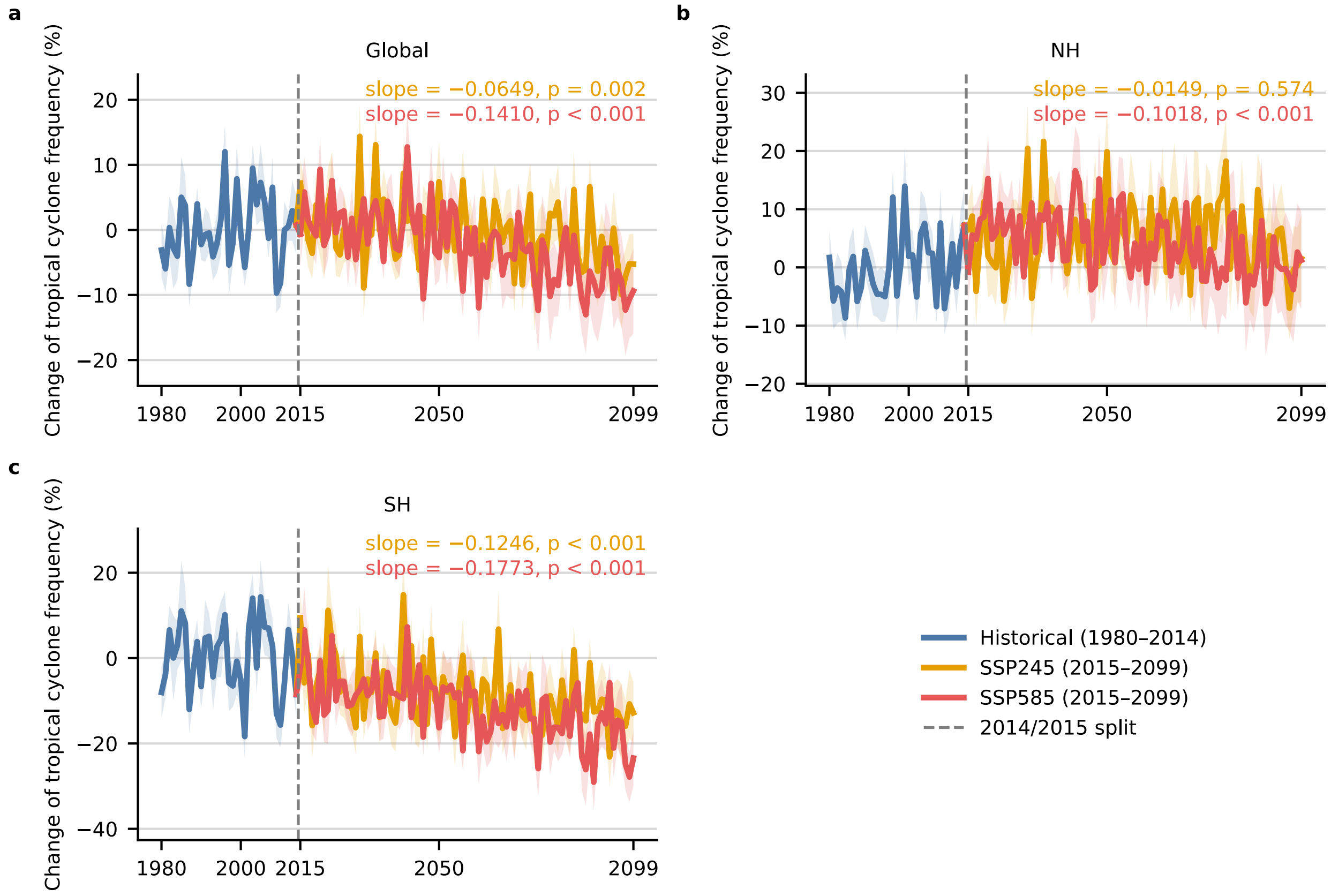

**Fig. 2 | Time series of multi-model mean changes in tropical cyclone frequency under SSP245 and SSP585 over 2015–2099.** **a**, Globe. **b**, Northern Hemisphere. **c**, Southern Hemisphere. Annual changes in tropical cyclone frequency relative to each model's present-day (1980–2014) climatology under SSP245 and SSP585 during 2015–2099, shown as multi-model means across the 14 CMIP6 models. Shading indicates ±1 standard error of the multi-model mean across the 14 CMIP6 models. The slope and p value shown in yellow for SSP245 and in red for SSP585 are derived from linear regression fitted to the corresponding multi-model-mean time series over 2015–2099.

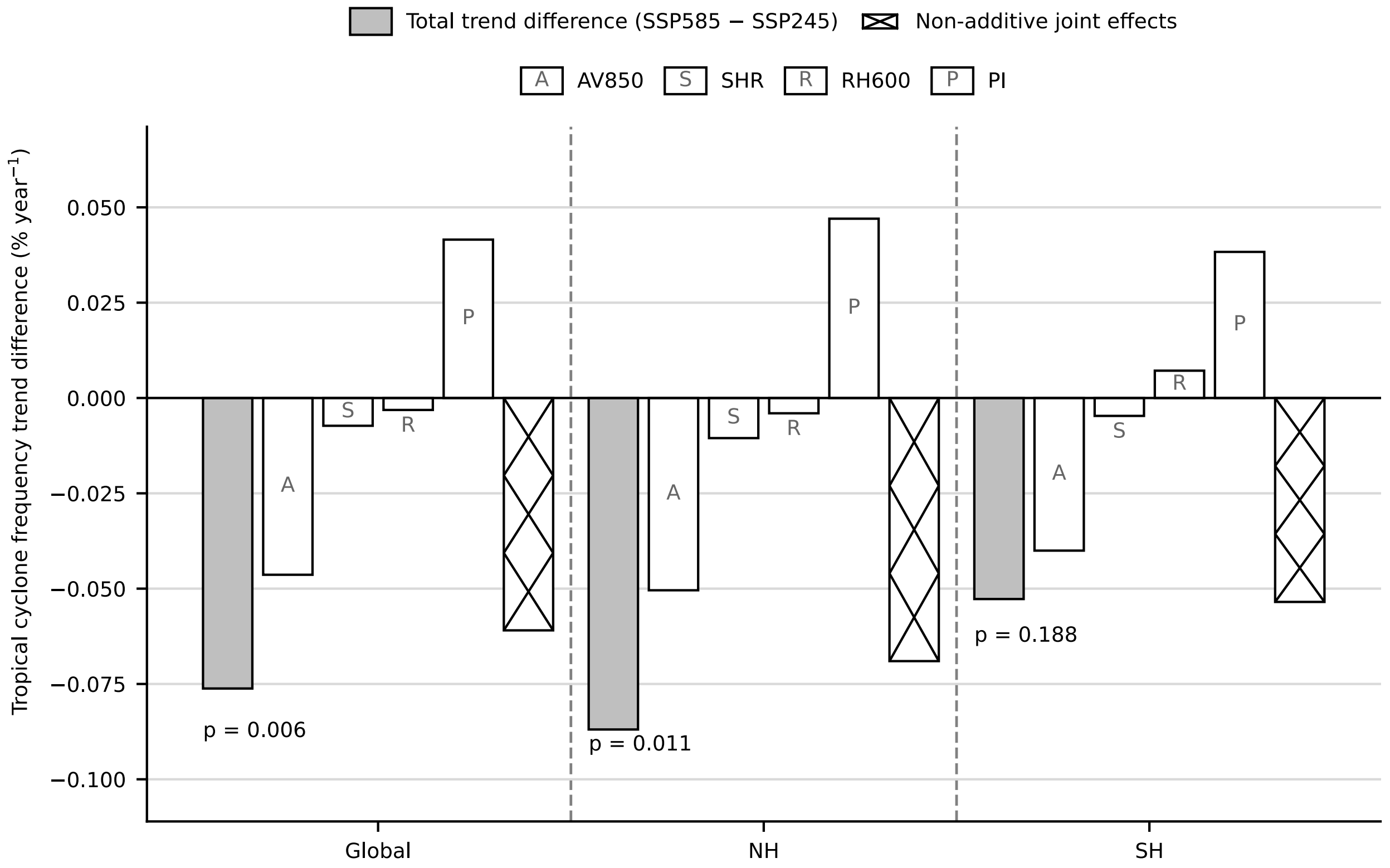

**Fig. 3 | Difference in the 2015–2099 tropical cyclone frequency trend between SSP585 and SSP245 decomposed into individual predictor contributions and non-additive joint effects.** Total difference in the linear trend of tropical cyclone frequency between SSP585 and SSP245 over 2015–2099 for the globe, Northern Hemisphere and Southern Hemisphere (see Methods). Grey bars show the total trend difference, bars labelled A, S, R and P show contributions from absolute vorticity at 850 hPa (AV850), vertical wind shear (SHR), relative humidity at 600 hPa (RH600), and potential intensity (PI); hatched bars show the contribution from non-additive joint effects. P values indicate the significance of the total trend difference (see Methods).

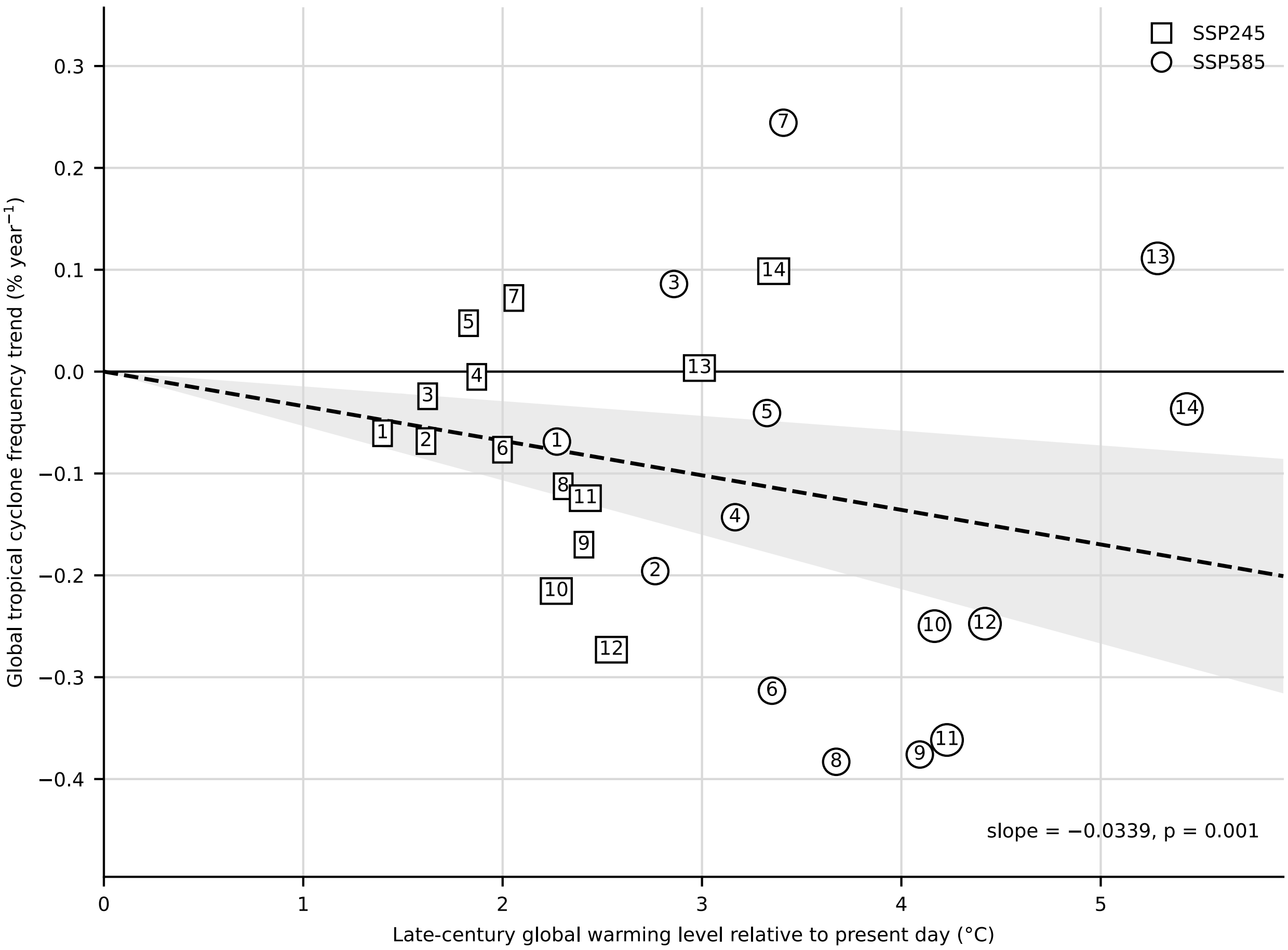

**Fig. 4 | Relationship between late-century global warming and the 2015–2099 trend in global tropical cyclone frequency across 28 CMIP6 model projections.** Late-century global warming level is global mean surface temperature over 2065–2099 relative to 1980–2014. Squares and circles denote SSP245 and SSP585 simulations, respectively. The dashed line and shaded region show the zero-intercept ordinary least-squares regression and 95% confidence interval (slope = $-0.0339$ % year$^{-1}$ °C$^{-1}$, $p = 0.001$). Numbers identify individual CMIP6 models and are the same as those shown in Extended Data Fig. 1.

## METHODS

### Data

We analysed monthly output from 14 selected CMIP6 models[36] for the historical experiment and the future scenarios SSP245 and SSP585. These models were chosen to span a broad range of projected late-century global warming levels (Extended Data Fig. 1). The historical simulations were used to define the present-day (1980–2014) climatology. Model names and CMIP6 variant labels are listed in Extended Data Table 1.

### PepC-Global genesis module

The PepC-Global[35] model is a basin-tuned framework that represents tropical cyclone climatology as a coupled stochastic process linking genesis, track and intensity conditioned on large-scale environmental predictors. In this study, we use only its genesis module. Genesis is represented as a probabilistic count process conditioned on large-scale environmental predictors on a 2.5° × 2.5° grid. The four large-scale environmental predictors are 850-hPa absolute vorticity, vertical wind shear between 850 and 200 hPa, relative humidity at 600 hPa, and potential intensity, representing two dynamical and two thermodynamic controls on genesis. Monthly genesis occurrence is treated as a Poisson process, but its rate parameter is estimated using support vector classification rather than conventional Poisson linear regression.

The support vector classifier uses a Gaussian radial basis function kernel, and the classifier output is treated as the Poisson rate parameter from which genesis events are sampled. Compared with Poisson linear regression, this formulation achieves higher out-of-sample skill across most basins and better reproduces basin-wise frequency, seasonal cycle, interannual variability and spatial distribution of genesis. A practical advantage of the support vector classification framework is that it can represent effectively zero genesis rates in strongly suppressed environments while still capturing sharp transitions in genesis favourability that are difficult for smooth log-linear formulations. This design is intended to better capture nonlinear, threshold-like environmental controls on genesis. For more details of the model, refer to ref. 35. In this study, for each CMIP6 projection we generate 100 stochastic realizations of tropical cyclone genesis.

**Counterfactual decomposition of trend differences between SSP585 and SSP245**

To quantify how the four genesis predictors contribute to difference in projected tropical cyclone frequency trends between SSP585 and SSP245, we used the PepC-Global genesis module to perform 16 counterfactual experiments for each scenario, corresponding to all possible subsets of the four predictors. For each subset $S$ of the four predictors $P = \{p_1, p_2, p_3, p_4\}$, the predictors included in $S$ retained their original monthly values during 2015–2099, whereas those excluded from $S$ were replaced, at each grid cell, by their 2015–2099 monthly climatological seasonal cycle under the same scenario. For each CMIP6 model and each counterfactual experiment, percentage changes were computed relative to that model's present-day (1980–2014) climatology, and the 2015–2099 linear trend of the percentage-change time series averaged across the 14 models is used as the response metric.

Let $\Delta v(S)$ denote the difference in tropical cyclone frequency trend between SSP585 and SSP245 obtained from subset $S$ experiments. It is estimated from the joint regression

$$Y_S(t,q) = \alpha + \beta_1 t + \beta_2 D_q + \Delta v(S) t D_q + \epsilon,$$

where $t$ is time, $q$ is scenario label, $Y_S(t,q)$ is the multi-model mean change of tropical cyclone frequency (%) from subset $S$ experiments, and $D_q$ is scenario indicator with 0 for SSP245 and 1 for SSP585. The p-value for $\Delta v(S)$ is computed from a two-sided $t$-test.

The individual contribution of predictor $p_i$ is estimated with all other predictors held at their monthly climatological seasonal cycles and is defined as

$$C_{p_i} = \Delta v(\{p_i\}) - \Delta v(\varnothing),$$

where $\Delta v(\varnothing)$ is the trend difference between SSP585 and SSP245 when all four predictors are fixed to their monthly climatological seasonal cycles, in which case neither scenario has a trend and thus $\Delta v(\varnothing) = 0$.

For any two-predictor subset $A \subset P$, the two-way non-additive joint effect is defined as

$$J(A) = \Delta v(A) - \sum_{\substack{S \subset A \\ |S|=1}} \Delta v(S), |A| = 2.$$

For any three-predictor subset $A \subset P$, the three-way non-additive joint effect is defined as

$$J(A) = \Delta v(A) - \sum_{\substack{S \subset A \\ |S|=2}} \Delta v(S) + \sum_{\substack{S \subset A \\ |S|=1}} \Delta v(S), |A| = 3.$$

The four-way non-additive joint effect among all four predictors is computed as

$$J(P) = \Delta v(P) - \sum_{\substack{S \subset P \\ |S|=3}} \Delta v(S) + \sum_{\substack{S \subset P \\ |S|=2}} \Delta v(S) - \sum_{\substack{S \subset P \\ |S|=1}} \Delta v(S).$$

Accordingly, the total difference in projected trend satisfies

$$\Delta v(P) = \sum_{p_i \in P} C_{p_i} + J_{\text{total}},$$

where $J_{\text{total}}$ is the total contribution from non-additive joint effects defined as

$$J_{\text{total}} = \sum_{\substack{S \subseteq P \\ |S| \geq 2}} J(S).$$

## DATA AVAILABILITY

All data used in this study are publicly accessible. CMIP6 model output is available through the Earth System Grid Federation at the CMIP6 archive portals (https://esgf-node.ornl.gov/search).

The data underlying Figs. 1–4 and Extended Data Figs. 1–2, including the derived tropical cyclone genesis frequency time series, counterfactual experiment outputs and global warming levels for each model–scenario combination, are archived on Zenodo[41].

## ACKNOWLEDGEMENTS

We acknowledge the World Climate Research Programme, the Working Group on Coupled Modelling, and the climate modelling groups participating in CMIP6 for producing and making available the model output used in this study. This work is supported by the National Science Foundation as part of the Megalopolitan Coastal Transformation Hub (MACH) under NSF award ICER-2103754. This is MACH contribution number 105. This work is also supported by the National Oceanic and Atmospheric Administration, US Department of Commerce, under award NA23OAR4320198, at Princeton University. Cong Gao is additionally supported by a postdoctoral fellowship from the Gordon and Betty Moore Foundation through Princeton University.

## AUTHOR CONTRIBUTIONS

C.G. and N.L. designed the research. C.G. performed the analysis and drew all the figures. C.G. wrote the first draft of the paper. All authors provided comments on different versions of the paper.

## COMPETING INTERESTS

The authors declare no competing interests.

## Extended Data Figures 1–2 and Extended Data Table 1

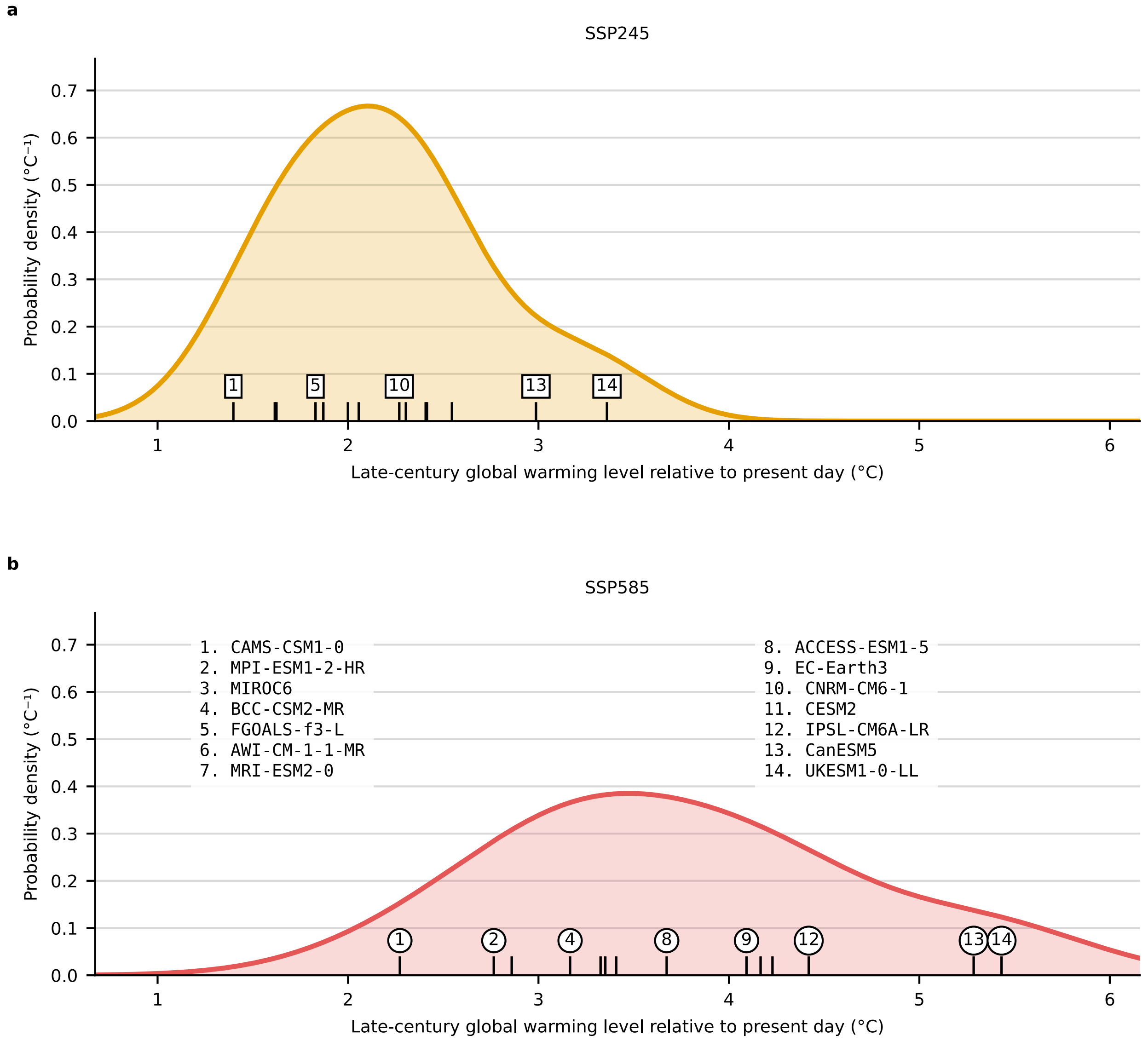

**Extended Data Fig. 1 | Probability density distributions of projected late-century global warming level under SSP245 and SSP585 across CMIP6 models. a**, SSP245. **b**, SSP585. Shaded curves show the distribution of global mean surface temperature change (2065–2099 relative to 1980–2014) across all 28 CMIP6 projections. Numbered ticks mark the 14 selected models used in this study (Extended Data Table 1).

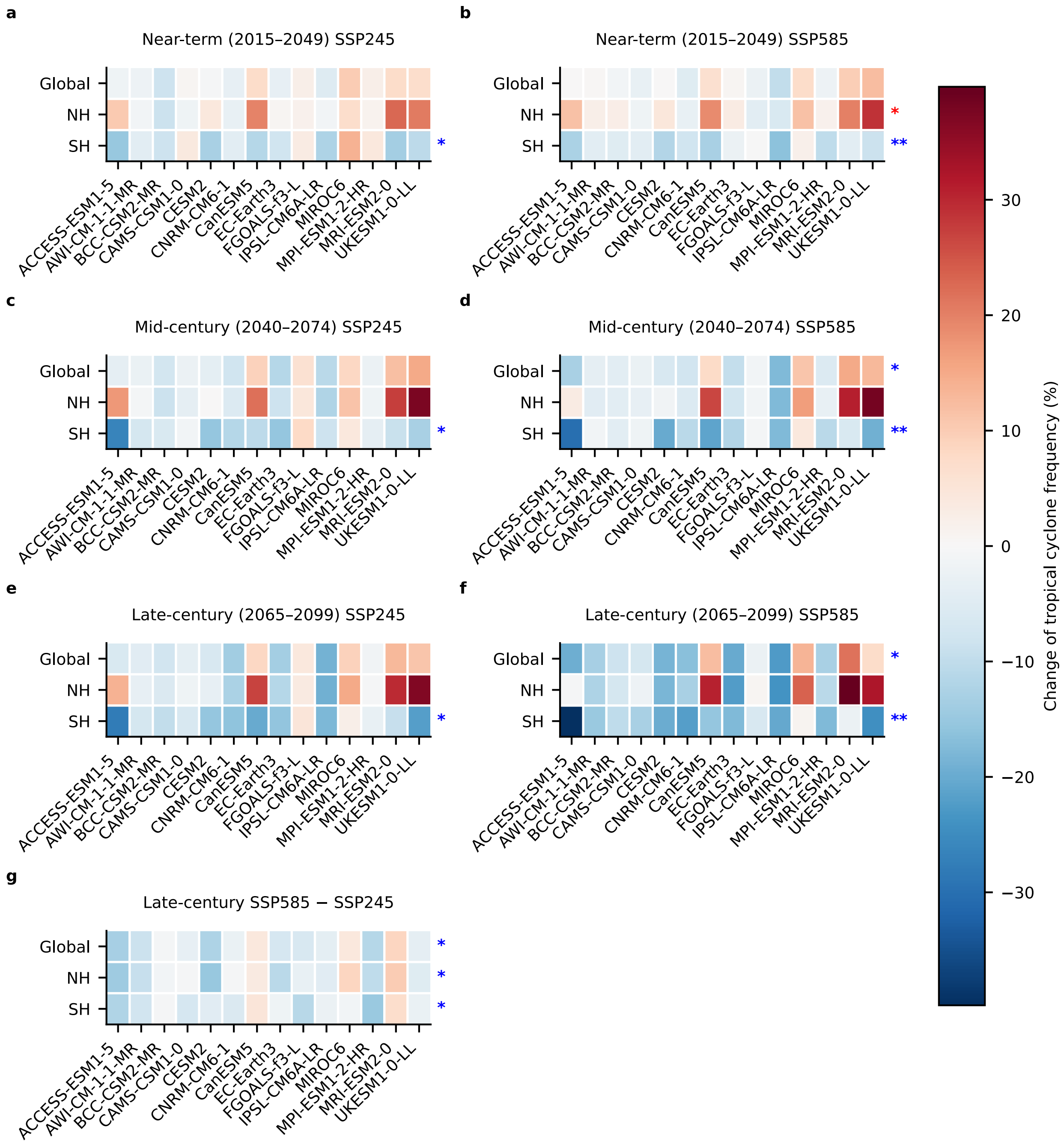

**Extended Data Fig. 2 | Projected changes in tropical cyclone frequency for individual CMIP6 models under SSP245 and SSP585.** **a**, Near-term (2015–2049) changes under SSP245. **b**, Near-term changes under SSP585. **c**, Mid-century (2040–2074) changes under SSP245. **d**, Mid-century changes under SSP585. **e**, Late-century (2065–2099) changes under SSP245. **f**, Late-century changes under SSP585. **g**, Late-century differences between SSP585 and SSP245. Changes are relative to the each model's present-day (1980–2014) climatology and are shown for the globe, Northern Hemisphere and Southern Hemisphere. Blue asterisks indicate consensus on negative changes and red asterisks indicate consensus on positive changes. One asterisk denotes likely agreement across models, defined as more than 66% of models showing the same sign of change,

and two asterisks denote very likely agreement, defined as more than 90% of models showing the same sign of change, following the calibrated uncertainty language of IPCC AR6.

**Extended Data Table 1 | CMIP6 models and variants used in this study.** For each model, r1i1p1f1 is used as the default variant when available, with r1i1p1f2 selected otherwise. For CESM2, however, r10i1p1f1 is used for the SSP245 and SSP585 simulations, because the original SSP r1i1p1f1 runs were retracted owing to a forcing-data bug and rerun with corrected forcing under new variant labels.

| Model | Historical variant | SSP245 variant | SSP585 variant |
|---|---|---|---|
| ACCESS-ESM1-5 | r1i1p1f1 | r1i1p1f1 | r1i1p1f1 |
| AWI-CM-1-1-MR | r1i1p1f1 | r1i1p1f1 | r1i1p1f1 |
| BCC-CSM2-MR | r1i1p1f1 | r1i1p1f1 | r1i1p1f1 |
| CAMS-CSM1-0 | r1i1p1f1 | r1i1p1f1 | r1i1p1f1 |
| CanESM5 | r1i1p1f1 | r1i1p1f1 | r1i1p1f1 |
| CESM2 | r1i1p1f1 | r10i1p1f1 | r10i1p1f1 |
| CNRM-CM6-1 | r1i1p1f2 | r1i1p1f2 | r1i1p1f2 |
| EC-Earth3 | r1i1p1f1 | r1i1p1f1 | r1i1p1f1 |
| FGOALS-f3-L | r1i1p1f1 | r1i1p1f1 | r1i1p1f1 |
| IPSL-CM6A-LR | r1i1p1f1 | r1i1p1f1 | r1i1p1f1 |
| MIROC6 | r1i1p1f1 | r1i1p1f1 | r1i1p1f1 |
| MPI-ESM1-2-HR | r1i1p1f1 | r1i1p1f1 | r1i1p1f1 |
| MRI-ESM2-0 | r1i1p1f1 | r1i1p1f1 | r1i1p1f1 |
| UKESM1-0-LL | r1i1p1f2 | r1i1p1f2 | r1i1p1f2 |